# A Beam Driven Plasma-Wakefield Linear Collider: From Higgs Factory to Multi-TeV

Summarized for CSS2013

E. Adli, J.P.Delahaye, S.J.Gessner, M.J. Hogan, T. Raubenheimer (SLAC)
W.An, C. Joshi, W.Mori (UCLA)

## 1. Introduction

Plasma wakefield acceleration (PWFA) holds much promise for advancing the energy frontier because it can potentially provide a 1000-fold or more increase in acceleration gradient with excellent power efficiency with respect to standard technologies. Most of the advances in beam-driven plasma wakefield acceleration were obtained by a UCLA/USC/SLAC collaboration working at the SLAC FFTB[i]. These experiments have shown that plasmas can accelerate and focus both electron and positron high energy beams, and an accelerating gradient in excess of 50 GV/m can be sustained in an 85 cm-long plasma. The FFTB experiments were essentially proof-of-principle experiments that showed the great potential of plasma accelerators.

The FACET[ii] test facility at SLAC will, in the period 2012-2016, further study several issues that are directly related to the applicability of PWFA to a high-energy collider, in particular two-beam acceleration where the witness beam experiences high beam loading (required for high efficiency), small energy spread and small emittance dilution (required to achieve luminosity).

The PWFA-LC concept presented in this document is an attempt to find a reasonable design that takes advantage of the PWFA, identify the critical parameters to be achieved and eventually the necessary R&D to address their feasibility. The design benefits from the extensive R&D that has been performed for conventional rf linear colliders during the last twenty years, especially ILC[iii] and CLIC[iv], with a potential for a substantially lower power consumption and cost.

## 2. A Plasma Wakefield Accelerator based Linear Collider

A novel design of a beam-driven PWFA linear collider is presented with geometric accelerating gradient on the order of 1 GV/m and extendable to the multi-TeV colliding beam energy range. The acceleration in plasma, being a single bunch process, provides great flexibility in the interval between bunches. In the preferred scheme, the main bunches collide in a continuous mode at several kHz repetition frequency. They are accelerated and focused with multi-GV/m fields generated in plasma cells powered by drive bunches with excellent transfer efficiency. The drive bunches are themselves accelerated by a CW superconducting rf recirculating linac, taking advantage of the impressive progress in the RF technology developed by ILC and providing excellent power efficiency together with high flexibility in the number of bunches. We consider the overall optimization for the proposed design, compare the efficiency with similar colliders like ILC and CLIC and outline the major R&D challenges. In a pulsed mode, the PWFA scheme could be used to upgrade a facility initially built with ILC technology up to the multi-TeV energy range.

### 3. A Conceptual PWFA Linear Collider

The concept for a PWFA-based Linear Collider is shown schematically in Figure 1 and the key parameters are provided in Table 1. It assumes similar processes for electron and positron acceleration although they could possibly be very different. Our approach uses established concepts for the particle and drive beam generation and focusing systems. However, this imposes important constraints on the plasma acceleration systems such as the need for high beam power and efficiency that are necessary for a realistic high energy linear collider design; many of these constraints are summarized in Ref. [v]. The current concept constitutes the basis for designing and understanding the proposed plasma wakefield research program at FACET while acknowledging that the detailed concepts for a PWFA-LC will continue to evolve with further study and simulation [vi].

**Figure 1: Layout of a 1 TeV PWFA Linear Collider**

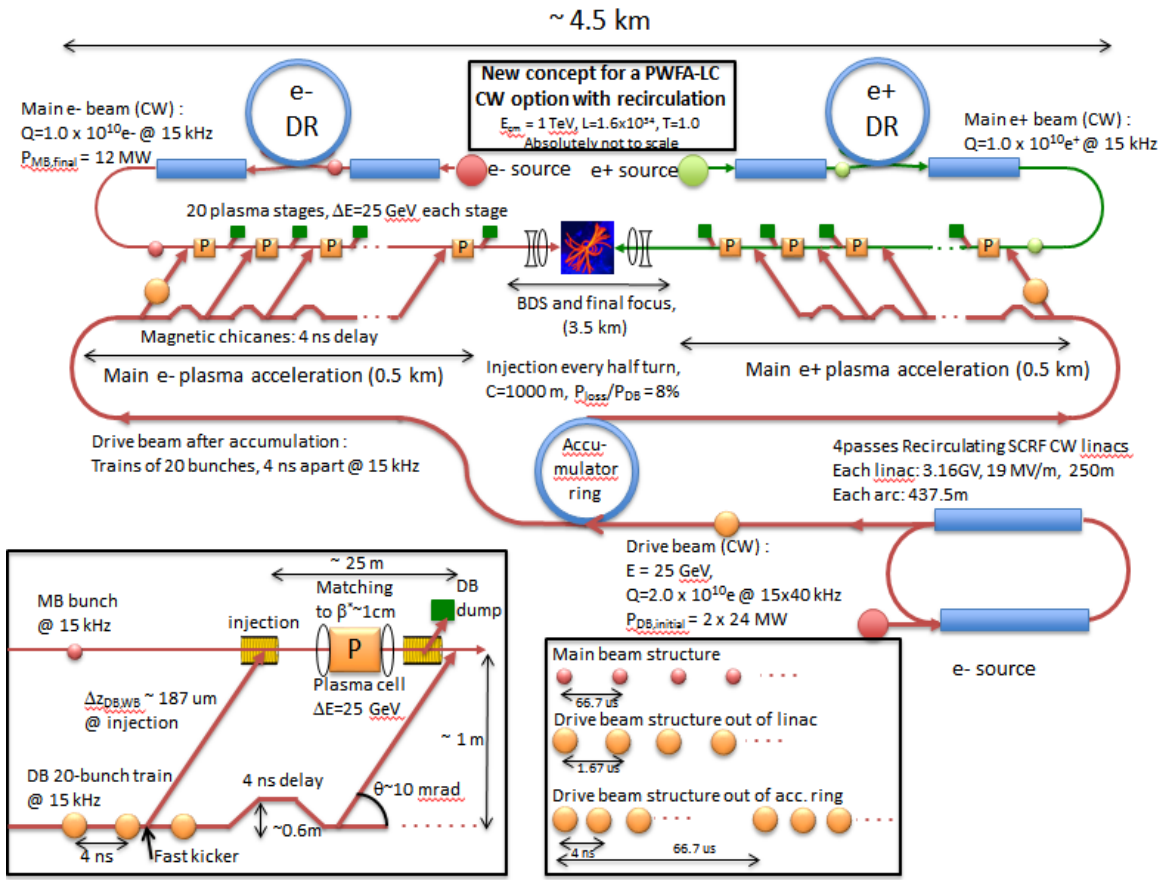

# Table 1: Main parameters at various beam collision energies

| E at IP, CM | GeV | 250 | 500 | 1000 | 3000 | 6000 | 10000 |
|---|---|---|---|---|---|---|---|
| N, experimental bunch | | 1.0E+10 | 1E+10 | 1.0E+10 | 1.0E+10 | 1.0E+10 | 1.0E+10 |
| Main beam bunches / train | | 1 | 1 | 1 | 1 | 1 | 1 |
| Main beam bunch spacing, | nsec | 3.33E+04 | 5.00E+04 | 6.67E+04 | 1.00E+05 | 1.43E+05 | 2.00E+05 |
| Repetition rate, | Hz | 30000 | 20000 | 15000 | 10000 | 7000 | 5000 |
| n exp.bunch/sec, | Hz | 30000 | 20000 | 15000 | 10000 | 7000 | 5000 |
| Avg current in exp beam | uA | 48.06 | 32.04 | 24.03 | 16.02 | 11.21 | 8.01 |
| peak current in exp beam | A | 4.81E-05 | 3.20E-05 | 2.40E-05 | 1.60E-05 | 1.12E-05 | 8.01E-06 |
| Power in exp. beam | W | 6.0E+06 | 8.0E+06 | 1.2E+07 | 2.4E+07 | 3.4E+07 | 4.0E+07 |
| Effective accelerating gradient | MV/m | 1000.00 | 1000.00 | 1000.00 | 1000.00 | 1000.00 | 1000.00 |
| Overall length of each linac | m | 125 | 250 | 500 | 1500 | 3000 | 5000 |
| BDS (both sides) | km | 2.00 | 2.50 | 3.50 | 5.00 | 6.50 | 8.00 |
| Overall facility length | km | 2.25 | 3.00 | 4.50 | 8.00 | 12.50 | 18.00 |
| | | | | | | | |
| **IP Parameters** | | | | | | | |
| Exp. bunch gamepsX, | m | 1.00E-05 | 1.00E-05 | 1.00E-05 | 1.00E-05 | 1.00E-05 | 1.00E-05 |
| Exp. bunch gamepsY, | m | 3.50E-08 | 3.50E-08 | 3.50E-08 | 3.50E-08 | 3.50E-08 | 3.50E-08 |
| beta-x, | m | 1.10E-02 | 1.10E-02 | 1.10E-02 | 1.10E-02 | 1.10E-02 | 1.10E-02 |
| beta-y, | m | 1.00E-04 | 1.00E-04 | 1.00E-04 | 1.00E-04 | 1.00E-04 | 1.00E-04 |
| sigx, | m | 6.71E-07 | 4.74E-07 | 3.35E-07 | 1.94E-07 | 1.37E-07 | 1.06E-07 |
| sigy, | m | 3.78E-09 | 2.67E-09 | 1.89E-09 | 1.09E-09 | 7.72E-10 | 5.98E-10 |
| sigz, | m | 2.00E-05 | 2.00E-05 | 2.00E-05 | 2.00E-05 | 2.00E-05 | 2.00E-05 |
| Y | | 8.44E-02 | 2.39E-01 | 6.75E-01 | 3.51E+00 | 9.93E+00 | 2.14E+01 |
| Dx | | 1.03E-02 | 1.03E-02 | 1.03E-02 | 1.03E-02 | 1.03E-02 | 1.03E-02 |
| Dy | | 1.83E+00 | 1.83E+00 | 1.83E+00 | 1.83E+00 | 1.83E+00 | 1.83E+00 |
| Uave | | 0.17 | 0.48 | 1.35 | 7.00 | 19.79 | 42.59 |
| delta_B | % | 2.75 | 6.66 | 12.76 | 23.10 | 27.67 | 29.88 |
| P_Beamstrahlung [W] | W | 1.7E+05 | 5.3E+05 | 1.5E+06 | 5.6E+06 | 9.3E+06 | 1.2E+07 |
| ngamma | | 0.57 | 0.73 | 0.88 | 1.05 | 1.11 | 1.14 |
| Hdx | | 1.00 | 1.00 | 1.00 | 1.00 | 1.00 | 1.00 |
| Hdy | | 4.62 | 4.62 | 4.62 | 4.62 | 4.62 | 4.62 |
| Hd | | 1.7 | 1.7 | 1.7 | 1.7 | 1.7 | 1.7 |
| **Geometric Lum (cm-2 s-1)** | | 9.41E+33 | 1.25E+34 | 1.88E+34 | 3.76E+34 | 5.27E+34 | 6.27E+34 |
| **Total Luminosity (cm-2 s-1)** | | 1.57E+34 | 2.09E+34 | 3.14E+34 | 6.27E+34 | 8.78E+34 | 1.05E+35 |
| **Integrated Lum. (fb-1 per 1E7s)** | | 157 | 209 | 314 | 627 | 878 | 1045 |
| Lum1% | | 9.41E+33 | 1.15E+34 | 1.57E+34 | 2.51E+34 | 3.07E+34 | 3.14E+34 |

The overall layout of a 1 TeV PWFA based Linear Collider shown in Figure 1 has a length of 4.5 km dominated by the final focus and beam delivery. The primary elements are two linacs with a length of 500 m each for 500 GeV acceleration with an effective field of 1 GV/m and a final focus and beam delivery system of 1.75km per side. The linacs are installed in a single tunnel without any active components except for the drive bunch distribution kickers. Each linac is made of 20 Two-Beam modules, 25 m long, which contain:

- One main line equipped with one high-gradient high-efficiency plasma cell and an interspace between plasma cells including matching optics, injection & extraction systems and beam instrumentation,
- A drive line including beam transfer optics and delay chicanes
- Transfer lines from drive line to main line
- Drive beam extraction from main line to a dump

The main beam and drive beam complex are located in a central position close to the detector with transport lines taking the beams to the beginning of the linacs. A more detailed description of the individual components is given below.

### 3.1. Main beam parameters and Luminosity

For comparison with existing designs of Linear Colliders like ILC and CLIC, the PWFA-LC is designed to be built in stages ranging over a broad beam collision energy range starting with a first stage at 250 GeV as a Higgs factory, upgradable to 500 GeV, 1 TeV, 3 TeV, 6 TeV and up to 10TeV. The main parameters including the total luminosity and luminosity in 1% of the peak energy are presented at the various beam collision energies. The PWFA luminosity and its comparison with the other Linear Colliders projects is displayed in Figure 2.

### 3.2. Main Beam Time Structure

Since the PWFA process is a single bunch acceleration process, the interval between bunches of the main beam can be chosen freely. An interval longer than 25ns is desirable to avoid pile-up of events in the detector and to make IP stabilization feedback more efficient. The ILC at 500 GeV has a bunch interval of 554 ns which is convenient for detector and beam stability considerations. At the other extreme, single bunch CW operation with a large interval between bunches and high bunch repetition frequency can mitigate pulsed and peak power to reduce cost and improve efficiency.

**Figure 2: PWFA luminosity and comparison with other Linear Collider options**

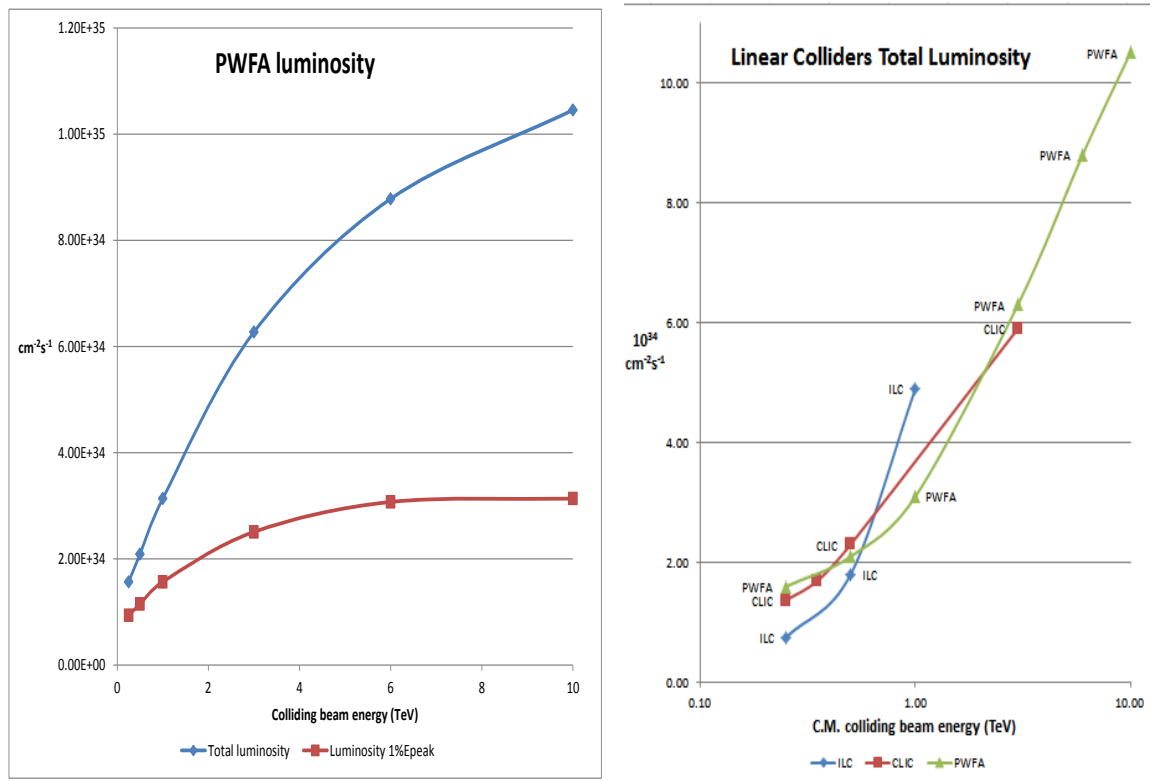

### 3.3. 500 GeV Design

The main beam parameters have been chosen to as close as possible to ILC at its nominal energy of 500 GeV. In particular, the beam transverse emittances take advantage of the design and R&D effort already performed for ILC. The only different parameters are:

- Charge per bunch of $1 \times 10^{10}$ as in present plasma tests instead of $2 \times 10^{10}$ in ILC.
- Bunch length of 20μm imposed by the plasma constraints instead of 300μm in ILC
- 20,000 bunches per second instead of 12,500 in ILC in order to partially compensate for the luminosity reduction due to the lower charge per bunch.
- Vertical focusing of the beam at the IP to $\beta_y$ = 0.1 mm as in CLIC instead of 0.48 mm in ILC, taking advantage of the reduced charge per bunch. As a consequence the vertical beam size at IP is reduced from 5.9 to 2.7 nm.

Finally, both the total luminosity and the luminosity within 1% of the peak energy, respectively $2.09 \times 10^{34}$ cm$^{-2}$s$^{-1}$ and $1.15 \times 10^{34}$ cm$^{-2}$s$^{-1}$, are slightly larger than for ILC in spite of a somewhat larger beamstrahlung ($\delta_b$ = 0.07 instead of 0.04) but still in the low beamstrahlung regime (Y =0. 24).

### 3.4. A First Stage for a Higgs Factory at 250 GeV

Similar parameters to the 500GeV design are adopted except for the number of bunches per second which is increased to 30,000 limited by the power consumption. The total luminosity and luminosity within 1% of the peak energy are respectively $1.6 \times 10^{34}$ cm$^{-2}$s$^{-1}$ and $0.94 \times 10^{34}$ cm$^{-2}$s$^{-1}$ in a low beamstrahlung regime (Y=0.03).

### 3.5. Energy Upgrade to the TeV and multi-TeV energy range

Similar parameters as for the 500 GeV design are adopted for the multi-TeV cases except for a reduced number of bunches per second in order to limit the power consumption. Because of the high beam collision energy, especially at 10 TeV, the beamstrahlung increases into the high-beamstrahlung regime (Y=21) but with a momentum spread $\delta_b$ = 0.30) much lower than at CLIC at 3 Tev ($\delta_b$ = 0.43) thanks to the smaller bunch length. Indeed as shown in [vii], the momentum spread scales with the square root of the bunch length in the high beamstrahlung regime (Y>>1).

### 3.6. Main Beam and Drive Beam Parameters:

The charge per bunch of the main and drive beams has been optimized for high acceleration field and efficient power transfer from drive to main beam resulting in $1 \times 10^{10}$ particles per bunch of the main beam and $2 \times 10^{10}$ e-/per bunch of the drive beam. Each stage has a plasma length of 3.3m at an accelerating gradient of 7.6 GeV/m to provide a main beam acceleration of 25 GeV, as described further in the plasma cell optimization description below.

The plasma cells are powered by trains of bunches produced using a Recirculating Linac Accelerator (RLA). Each drive bunch powers one single plasma cell accelerating one single main bunch by 25 GeV. The drive bunch is then dumped and a fresh drive bunch is used to power the following plasma cell for further acceleration of the main bunch. The process starts with a CW SC linac for optimum efficiency and a recirculating beamline to reduce the overall drive beam linac length and the associated cost and cryogenics power. The bunches are fed into an accumulator ring to generate the time structure required to power the plasma stages. When enough bunches to accelerate a single electron and positron bunch to final energy have been accumulated in the ring, they are extracted and distributed to the plasma cells from a co-linear distribution system. This system uses fast kickers, small angle bends and magnetic chicanes as delay lines to satisfy the required timing.

The basic concept of an injection/extraction system for a standard inter-plasma cell has been developed and fulfills four main functions: i) matching of the main beam to the entry of the next plasma cell, ii) extraction of the drive beam used in the previous plasma cell and transfer to a dump, iii) transfer of a fresh drive bunch from the drive beam line and iv) injection into the following plasma cell. These elements are described in detail in [viii].

## 4. Efficiency

The efficiency of any accelerator is critical but this is particularly true for a linear collider where the average power level in the output beams are tens of MW. In this section we discuss the efficiency of transferring the drive beam energy to a trailing beam. Obtaining high efficiency is challenging but it is made even more difficult when stringent conditions are placed on the beam quality of the trailing beam (small energy spread and emittance). To study the efficiency one can split the analysis into two parts. The first is the efficiency of transferring the drive beam energy into the wake and the second is transferring the energy in the wake into the trailing beam. The first analysis requires studying wakefield excitation and determining the distance over which the wake can be excited. For PWFA, the distance over which the wake can be excited is limited by drive beam depletion (and distortion) while dephasing between the drive and trailing beam and beam diffraction are generally not issues. The transferring of wake energy into the trailing beam is the topic of beam loading.

The topics of depletion, and beam loading all have been studied extensively using linear theory, nonlinear theory, and fully nonlinear, fully self-consistent particle-in-cell simulations.

In the nonlinear blowout regime, the lower bound on beam to wake efficiency gets significantly higher than the 50-60% values typically derived using linear theory. For example, simulations with a beam population of $3\times10^{10}$ electrons (a beam density of $3.6\times10^{17} cm^{-3}$) and the plasma density of $10^{16} cm^{-3}$, the peak decelerating field is at maximum close to the center of the beam and more importantly the decelerating field is constant in radius so that particles within a given slice all slow down together. As a result, 81% of the drive beam energy was transferred to the wake in this case.

The above simulations were for symmetric Gaussian beams and for such beams the wakes inside the beam are not constant so the beam does not slow down together. A higher efficiency can be obtained by tailoring the drive beam so that the decelerating field inside beam is constant. Early work in the linear regime [ix] showed that a wedge shaped (with a precursor) would lead to a uniform decelerating field and to a high transformer ratio. W. Lu et al. [x] showed that in the nonlinear blowout the optimum current profile is also a linear rise. Huang, Tzoufras et al, have shown that a precursor can help in the nonlinear regime as well. The point is that efficiencies in excess of 80% from the drive beam to the wake are feasible for unshaped electron bunches and greater than 90% for shaped bunches.

Obtaining a high beam loading efficiency [xi] is challenging and also requires shaping the trailing bunches or operating in the nonlinear regime.

In the linear regime beam loading is analyzed by calculating the wakes from the drive beam and the trailing beam independently and then using linear superposition. Knowing the combined wakes inside the bunch allows an analysis of energy spread and emittance. The beam loading efficiency can be obtained by comparing the wake in front and behind the trailing beam. If the wake behind the trailing beam is zero then 100% beam loading efficiency can be achieved. If the trailing beam is very short then 100% beam loading efficiency can be achieved but at the expense of 100% energy spread. Many options for beam loading and the tradeoffs in efficiency vs. wake amplitude were discussed in the work of Katsouleas et al. [xii]. However, as eluded to in that work, one option that required tightly focused trailing beams naturally led to the conclusion that the trailing beam itself will excite nonlinear wakes.

In 2008 Tzoufras developed a theoretical framework for studying beam loading in the nonlinear regime. There were several important conclusions. The first is that in the nonlinear regime (valid for accelerating electrons) the beam loading efficiency can also exceed 90% for shaped trailing bunches while maintaining low energy spread and emittance (assuming the ions don't move). The second is that high efficiencies and low energy spread can still be achieved for unshaped Gaussian bunches.

## 5. Plasma Cell Optimization

The present design uses an energy gain of 25 GeV per stage and 20 stages with a main beam bunch charge of $10^{10}$. The parameter optimization presented here assumes acceleration of an electron witness bunch driven by an electron drive bunch. The drive beam parameters are considered free variables, which can be chosen to minimize the main beam energy spread, maximize the drive beam to main beam efficiency, minimize power consumption and minimize the drive beam energy in order to ease the injection/extraction requirements. A minimum plasma density of $2 \times 10^{16}/cm^3$ is chosen for a gradient in the plasma of 7.6 GV/m resulting in an acceptable plasma cell length of 3.3 m for a 25 GeV energy gain per cell.

To minimize the energy spread, the witness bunch needs to load the accelerating wake correctly and create a constant accelerating field where the bulk of the main beam charge is. This can be achieved following the paper of Tzoufras on non-linear beam loading [xiii]. By adjusting the ratio of the witness bunch and drive bunch charges, for a given distance between the two bunches, the beam loading of the main bunch leads to a flat wake around its center. The distance between the two bunches decides the ratio of the flat part of the accelerating field versus the peak decelerating field; the transformer ratio.

In the blow-out regime, the peak accelerating field can be several times higher than the peak decelerating field, allowing for a transformer ratio larger than 1 as shown on [viii]. A transformer ratio larger than 1 reduces the required drive beam energy, but tightens the main bunch injection tolerances, as the witness bunch needs to be positioned relatively closer to the edge of the bubble. A transformer ratio of 1 has therefore been adopted.

The parameters have been optimized using non-linear theory as well as iterations with the code QuickPIC[xiv]. The length of the drive bunch is set as $\sim 1/k_p$ $\sim$40 um, the optimal bunch length to drive the wake. For the constraints described above, the required drive bunch charge is $2\times10^{10}$ e$^-$. The distance between the drive bunch and the witness bunch resulting in a transformer ratio of 1 is 187 um.

According to QuickPIC simulations, conservatively using non-shaped Gaussian beam current profiles, the resulting parameters yield an overall drive bunch to witness bunch power transfer efficiency of 50% resulting from a 77% drive to plasma transfer efficiency and a 65% plasma to witness bunch transfer efficiency [viii]. These simulations are performed with an e- drive bunch and an e- witness bunch, assuming ions don't move. For positron acceleration other regimes such as the near hollow channel proposed most recently by [xv] shows promise, however precise efficiency calculations have not yet been performed for this regime. The power flow is summarized in the figure 3 below:

**Figure 3: Power flow from an e- drive bunch (DB) to Plasma and to an e- Witness Bunch (WB), as calculated by QuickPIC.**

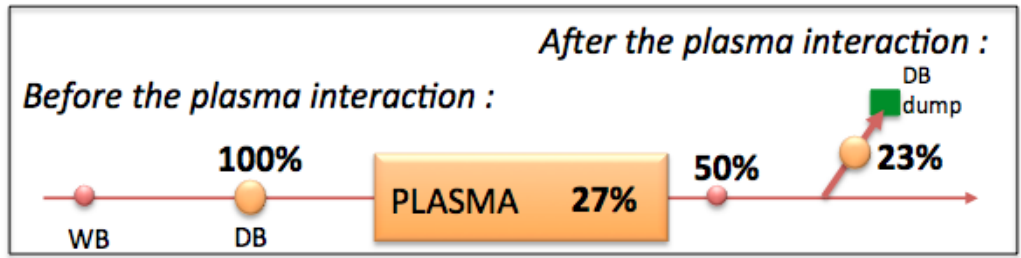

## 6. Wall Plug Power Estimation

An overall drive beam power and wall plug power consumption of the overall complex is estimated in Table 2 for various colliding beam energies.

It assumes realistic Klystron power efficiency of 90% and Klystron RF to beam efficiency of 65% based on LEP or CEBAF experience with CW operation. With these efficiencies, the RF power to accelerate the drive beam up to the requested energy of 25 GeV varies from 26 to 223 MW at 250 GeV and 10 TeV respectively. In addition 1 to 42 MW have to be provided to compensate for synchrotron radiation in the accumulator ring. That corresponds to about 27 to 265 klystrons of 1 MW each. Thus the wall plug power for drive beam acceleration varying from 61 to 416 MW corresponds to the lion's share of the total wall power consumption. The cryogenic power of the SC linacs is limited to 15.7 MW by recirculation in the CW drive linac.

A wall plug consumption ranging from 56 to 105 MW is added for the main and drive beam injector complexes based on detailed CLIC estimation. The main beam acceleration of about 20% is especially efficient thanks to the CW operation of the Superconducting drive linac and the excellent drive to beam transfer efficiency of the plasma. The corresponding wall plug to beam efficiency ranging from 9.1% at 250 GeV to 15.1% at 10 TeV are comparable to ILC at low energy and particularly attractive at high energy as compared with CLIC (4.8% at 3 TeV).

As a consequence, the wall plug consumption ranging from 133 MW at 250 GeV to 318 MW at 3 TeV and 537 MW at 10 TeV are comparable to ILC at low energy and lower by a factor two in respect with CLIC at the same energy. The figure of merit defined as the luminosity normalised to the overall wall plug power makes the PWFA a very attractive technology for high energy applications as shown on the figure 4.

**Table 2: Drive beam parameters and input beam power for plasma based linear collider design based on plasma cell transformer ratio of 1.**

| Case | | 250 | 500 | 1000 | 3000 | 6000 | 10000 |
|---|---|---|---|---|---|---|---|
| | GeV | | | | | | |
| Transfer efficiency drive to main | % | 50.00 | 50.00 | 50.00 | 50.00 | 50.00 | 50.00 |
| Number of plasma cell per linac | _ | 5.00 | 10.00 | 20.00 | 60.00 | 120.00 | 200.00 |
| Drive bunch repetition frequency in drive linac | Hz | 300000 | 400000 | 600000 | 1200000 | 1680000 | 2000000 |
| Interval between drive bunches in drive linac | m | 1000 | 750 | 500 | 250 | 179 | 150 |
| Drive linac intensity with 4 recirculations | mA | 3.89 | 5.18 | 7.77 | 15.54 | 21.76 | 25.90 |
| Drive beam power per drive beam | MW | 12.15 | 16.20 | 24.30 | 48.60 | 68.04 | 81.00 |
| Synchrotron energy lost in recirculation linac | GeV | 0.28 | 0.28 | 0.28 | 0.28 | 0.28 | 0.28 |
| Power lost in recirculation linac | MW | 1.10 | 1.47 | 2.20 | 4.41 | 6.17 | 7.35 |
| RF voltage in drive linac | GeV | 6.53 | 6.53 | 6.53 | 6.53 | 6.53 | 6.53 |
| Average beam current in Accumulator | ma | 4.86 | 9.71 | 19.43 | 58.28 | 116.57 | 194.28 |
| Synchrotron energy losses per turn Accumulator | GeV | 0.22 | 0.22 | 0.22 | 0.22 | 0.22 | 0.22 |
| Beam Power loss in Accumulator | MW | 1.06 | 2.11 | 4.22 | 12.67 | 25.35 | 42.25 |
| RF power for 2 drive beams acceleration | MW | 26.46 | 35.98 | 55.03 | 120.30 | 176.42 | 222.73 |
| Cryogenic power | MW | 15.72 | 15.72 | 15.72 | 15.72 | 15.72 | 15.72 |
| Wall plug power for drive beams acceleration | MW | 60.95 | 77.23 | 109.79 | 232.18 | 333.17 | 416.50 |
| Total wall plug power | MW | 132.92 | 150.45 | 185.51 | 317.90 | 433.89 | 537.22 |
| Drive beam acceleration efficiency (incl cryo) | % | 39.87 | 41.95 | 44.27 | 41.86 | 40.84 | 38.90 |
| Beam acceleration efficiency | % | 19.94 | 20.98 | 22.13 | 20.93 | 20.42 | 19.45 |
| Wall plug to main beam efficiency | % | 9.14 | 10.77 | 13.10 | 15.29 | 15.68 | 15.08 |
| Figure of merit: L/Wall plug power | L/MW | 7.08 | 7.64 | 8.45 | 7.89 | 7.08 | 5.84 |

**Figure 4: Wall plug power consumption and luminosity to power figure of merit in Linear Colliders**

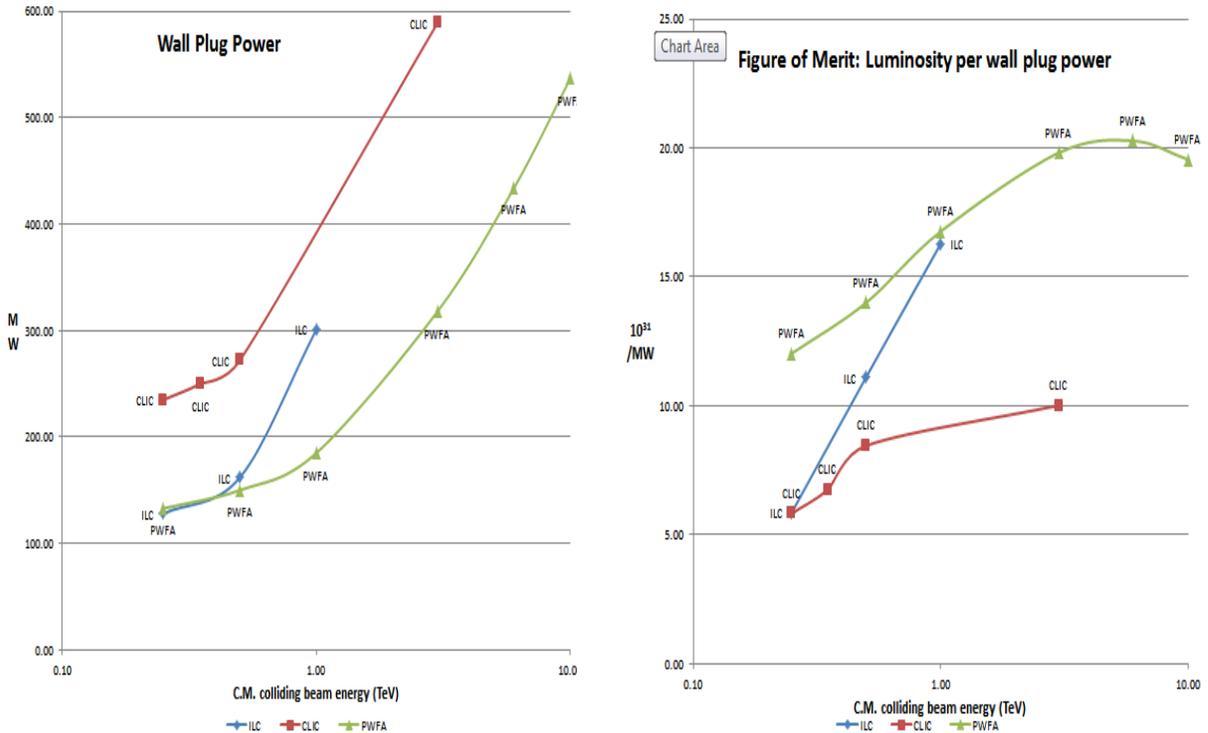

## 7. Pulsed operation mode as an alternative for ILC upgrade

Thanks to the flexibility of the interval between bunches, the PWFA technology can also be used in a pulsed mode to accelerate a beam with parameters and train structure very similar to the one of the ILC except for the bunch length which has to be reduced by a factor 15 from 300 to 20 microns. Assuming that the feasibility of the PWFA technology has been fully demonstrated by then, it could be considered as a possible alternative for an ILC energy upgrade above 250 GeV as envisaged for a HIGGS factory towards TeV and multi-TeV energy range with parameters summarized in table 3 below and without any modification of the ILC facility except for the implementation of a bunch compressor. With such an upgrade, the total extension of an ILC/PWFA complex at 1 TeV would be limited to 21 km as the ILC based Higgs facility instead of 52 km as presently foreseen with upgraded ILC technology.

Table 3: ILC energy upgrade from 250 GeV to 1 TeV by PWFA

| Parameter | Unit | ILC | ILC | ILC (to 250GeV) + PWFA (250to1000) | |
|---|---|---|---|---|---|
| | | | | High current (NC fully loaded) | Low current (SC 325MHz) |
| Energy (cm) | GeV | 250 | 1000 | | |
| Luminosity (per IP) | $10^{34} cm^{-2} s^{-1}$ | 0.75 | 4.9 | 4.9 | 4.1 |
| Peak (1%)Lum(/IP) | $10^{34} cm^{-2} s^{-1}$ | 0.65 | 2.2 | 2.2 | 2.5 |
| # IP | - | 1 | 1 | 1 | 1 |
| Length | km | 21 | 52 | 21 | 21 |
| Power (wall plug) | MW | 128 | 300 | 128+135*1.2=290? | 300? |
| Polarisation (e+/e-) | % | 80/30 | 80/30 | 80/30 | 80/30 |
| Lin. Acc. grad. (peak/eff) | MV/m | 31.5/25 | 36/30 | 7600/1000 | 7600/1000 |
| # particles/bunch | $10^{10}$ | 2 | 1.74 | 1.74 | 1.00 |
| # bunches/pulse | - | 1312 | 2450 | 2450 | 1312 |
| Bunch interval | ns | 554 | 366 | 366 | 554 |
| Average/peak current | nA/mA | 21/6 | 22.9/7.6 | 22.9/7.6 | 27/2.9 |
| Pulse repetition rate | Hz | 5 | 4 | 5 | 13 |
| Beam power/beam | MW | 2.63 | 13.8 | 13.8 | 14.0 |
| Drive beam pulsed current | mA | - | - | 600 | 231 |
| Norm Emitt (X/Y) | $10^{-6}/10^{-9}$ rad-m | 10/35 | 10/30 | 10/30 | 10/30 |
| Sx, Sy, Sz at IP | nm,nm,μm | 729/6.7/300 | 335/2.7/225 | 485/2.7/20 | 485/2.7/20 |
| Crossing angle | mrad | 14 | 14 | 14 | 14 |
| Av # photons | - | 1.17 | 2.0 | 1.0 | 0.9 |
| δb beam-beam | % | 0.95 | 10.5 | 16 | 13 |
| Upsilon | - | 0.02 | 0.09 | 0.8 | 0.7 |

Alternatively, the PWFA technology could also be used as an afterburner of the ILC:
After beam acceleration up to an initial energy with ILC technology, the beam could be further accelerated with PWFA technology at very low cost. Each ILC would be split in two bunches, one with 2/3 of the charge used as drive bunch and a second with 1/3 of the charge used as main bunch. By transfer of energy of the drive to the main bunches in the plasma cells, the ILC beam energy could be doubled without any drive beam injector complex and without any substantial additional power. As shown on table 4 comparing parameters of a 1 TeV collider based on pure ILC technology with the one based on PWFA used as ILC afterburner from 500 GeV, a similar luminosity can be obtained in spite of the low charge per bunch with stronger horizontal focusing at IP and higher repetition rate.

**Table 4: ILC energy upgrade from 500 GeV to 1 teV by PWFA afterburner**

| Parameter | Unit | ILC | ILC | ILC + PWFA |
|---|---|---|---|---|
| Energy (cm) | GeV | 500 | 1000 | PFWA = 500 to 1000 |
| Luminosity (per IP) | $10^{34} cm^{-2} s^{-1}$ | 1.5 | 4.9 | 2.6 |
| Peak (1%)Lum(/IP) | $10^{34} cm^{-2} s^{-1}$ | 0.88 | 2.2 | 1.3 |
| # IP | - | 1 | 1 | 1 |
| Length | km | 30 | 52 | 30 |
| Power (wall plug) | MW | 128 | 300 | 150? |
| Polarisation (e+/e-) | % | 80/30 | 80/30 | 80/30 |
| Lin. Acc. grad.(p/eff) | MV/m | 31.5/25 | 36/30 | 7600/1000 |
| # particles/bunch | $10^{10}$ | 2 | 1.74 | 0.66 |
| # bunches/pulse | - | 1312 | 2450 | 2450 |
| Bunch interval | ns | 554 | 366 | 366 |
| Averag/peak current | nA/mA | 21/6 | 22.9/7.6 | 22.9/7.6 |
| Pulse repetition rate | Hz | 5 | 4 | 15 |
| Beam power/beam | MW | 5.2 | 13.8 | 13.8 |
| Norm Emitt (X/Y) | $10^{-6}/10^{-9}$ radm | 10/35 | 10/30 | 10/30 |
| Sx, Sy, Sz at IP | nm,nm,μm | 474/5.9/300 | 335/2.7/225 | 286/2.7/20 |
| Crossing angle | mrad | 14 | 14 | 14 |
| Av # photons | - | 1.70 | 2.0 | 0.7 |
| δb beam-beam | % | 3.89 | 9.1 | 9.3 |
| Upsilon | - | 0.03 | 0.09 | 0.52 |

## 8. Technology applications

The concept described in this document is derived for High Energy applications, but PWFA technology may be used for other very attractive applications taking advantage of large accelerating beams in the plasma, especially:

1. Generation of beams with extremely small emittances, so-called Trojan horse technique.
2. A Compact X-FEL using the plasma as a high-gradient accelerator and a source of high-brightness beams (Figure 5)

**Figure 5: Layout of a PWFA based X-FEL**

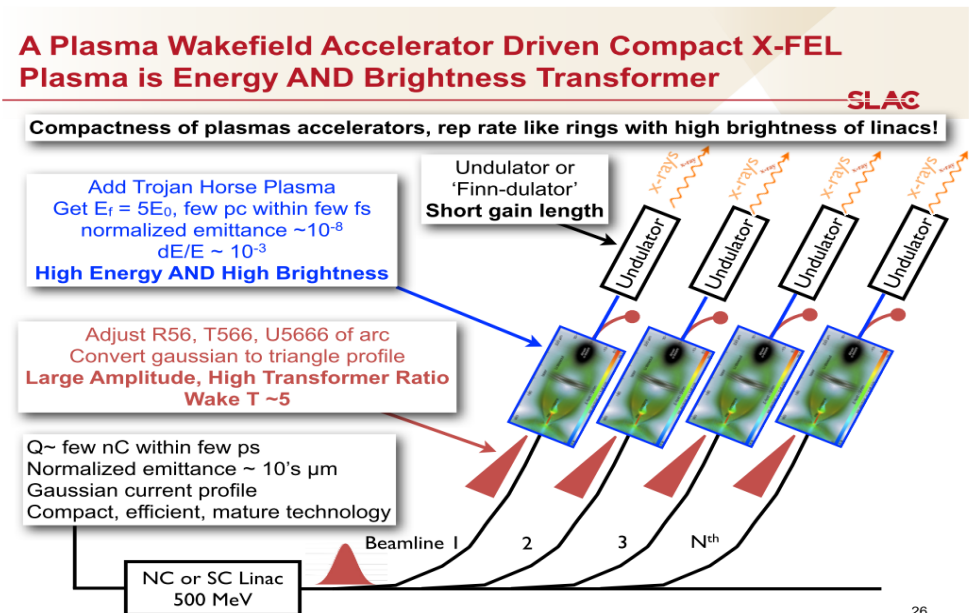

## 9. Primary Issues for a PWFA-LC

The concept for the PWFA-LC highlights the key beam and plasma physics with challenges which must be addressed by experimental facilities such as FACET. A reasonable set of design choices for a plasma-based linear collider can benefit from the years of extensive R&D performed for the beam generation and focusing subsystems of a conventional rf linear collider. The remaining experimental R&D is directly related to the beam acceleration mechanism. In particular, the primary issues are:

- Development of a concept for positron acceleration with high beam brightness
- High beam loading with both electrons and positrons (required for high efficiency),
- Beam acceleration with small energy spreads (required to achieve luminosity and luminosity spectrum),
- Preservation of small electron beam emittances (required to achieve luminosity) and mitigation of effects resulting from ion motion
- Preservation of small positron beam emittances (required to achieve luminosity) and mitigation of effects resulting from plasma electron collapse
- Average bunch repetition rates in the 10's of kHz (required to achieve luminosity)
- Synchronization of multiple plasma stages to achieve the desired energy, and
- Optical beam matching between plasma acceleration stages and from plasma to beam delivery systems.

## 10. R&D and tentative schedule

An ambitious test facility, FACET, operated as a user facility at SLAC and taking advantage of the dense electron and positron bunches provided by the 20 GeV linac is ideal to directly address with a targeted experimental program a number of critical issues listed above over the next four years. Desire to address the remaining issues has led to a concept for a follow on facility dedicated to studying beam-driven plasma wakefield acceleration called FACET-II aiming for a feasibility demonstration within a decade. An extensive design and simulation effort must proceed in parallel with the FACET experimental effort to both support the experimental program and to fully develop the PWFA-LC design concepts outlined here.

When the feasibility of the PWFA technology will have been demonstrated, it may be used for applications with strong physics interest and gradually increasing complexity in a staged approach like the one described below with a tentative schedule on Figure 6. That will help to further develop and validate the technology by integration of the various systems and to get operational experience which is necessary before a more complex or higher energy application can be envisioned.

- It could be used initially for attractive low energy applications taking advantage of large accelerating fields in the plasma, especially the generation of beams with extremely small emittances, so called Trojan horse technique and/or a Compact X-FEL using the plasma as a high-gradient accelerator and a source of high-brightness beams.
- A larger scale application for ILC energy upgrade could then be envisaged using initially the ILC (if operational at the time) as an R&D platform for further validation of the PWFA technology, possibly in parallel with ILC operation as a Higgs factory from 2025, starting with single stage evolving gradually towards multi–stages accelerations studies.
- If successful, the PWFA technology could be considered as a possible candidate for ILC energy upgrade from 2030 (if requested by Physics at the time).

**Figure 6: Tentative PWFA schedule for R&D and possible applications**

| Technological issues | | 2013 | 2014 | 2015 | 2016 | 2017 | 2018 | 2019 | 2020 | 2021 | 2022 | 2023 | 2024 | 2025 | 2026 | 2027 | 2028 | 2029 | 2030 | 2031 | 2032 | 2033 |
|---|---|---|---|---|---|---|---|---|---|---|---|---|---|---|---|---|---|---|---|---|---|---|
| Systems | Components & options | | | | | | | | | | | | | | | | | | | | | |
| Test facilities | FACET | ■ | ■ | ■ | ■ | ■ | | | | | | | | | | | | | | | | |
| | FACET II | | ■ | ■ | ■ | ■ | ■ | ■ | ■ | ■ | ■ | | | | | | | | | | | |
| | ILC as Higgs factory @ 250GeV | | | | ■ | ■ | ■ | ■ | ■ | ■ | ■ | ■ | ■ | ■ | ■ | ■ | ■ | ■ | | | | |
| | ILC as R&D platform | | | | | | | | | | | ■ | ■ | ■ | ■ | ■ | ■ | ■ | ■ | | | |
| Key issues | development of a concept for positron acceleration with high beam brightness | ■ | ■ | ■ | ■ | ■ | ■ | ■ | ■ | ■ | ■ | | | | | | | | | | | |
| | High beam loading with both electrons and positrons | ■ | ■ | ■ | ■ | ■ | ■ | ■ | ■ | ■ | ■ | | | | | | | | | | | |
| | Beam acceleration with small energy spreads | ■ | ■ | ■ | ■ | ■ | ■ | ■ | ■ | ■ | ■ | | | | | | | | | | | |
| | Preservation of small electron beam emittances and mitigation of effects resulting from ion motion | ■ | ■ | ■ | ■ | ■ | ■ | ■ | ■ | ■ | ■ | | | | | | | | | | | |
| | Positron beam emittances preservation and mitigation of effects resulting from plasma electron collapse | ■ | ■ | ■ | ■ | ■ | ■ | ■ | ■ | ■ | ■ | | | | | | | | | | | |
| | Average bunch repetition rates in the 10's of kHz | | | | | | ■ | ■ | ■ | ■ | ■ | | | | | | | | | | | |
| | Synchronization of multiple plasma stages | | | | | | ■ | ■ | ■ | ■ | ■ | | | | | | | | | | | |
| | Optical beam matching between plasma acceleration stages and from plasma to beam delivery systems. | | | | | | ■ | ■ | ■ | ■ | ■ | | | | | | | | | | | |
| Integrated systems with Physics applications | Beam generation with extremely small emittances (Trojan horse technique) | | | | | | ■ | ■ | ■ | ■ | ■ | ■ | ■ | ■ | ■ | ■ | ■ | ■ | ■ | ■ | ■ | ■ |
| | Compact X-FEL using the plasma as a high-gradient accelerator and a source of high-brightness beams.. | | | | | | ■ | ■ | ■ | ■ | ■ | ■ | ■ | ■ | ■ | ■ | ■ | ■ | ■ | ■ | ■ | ■ |
| | ILC energy upgrade | | | | | | | | | | | | | ■ | ■ | ■ | ■ | ■ | ■ | ■ | ■ | ■ |

Color code:
- R&D feasibility, Conceptual design (yellow)
- Technical design (red)
- Construction (green)
- Operation (blue)

## 11. Conclusion

Beam driven Plasma wake-Field Accelerators (PWFA) provides a very attractive novel technology with large accelerating fields and excellent power efficiency for a number of applications ranging from Photon science to High Energy Physics with a broad range of energy from HIGGS factory to Multi-TeV.

Many of the critical issues are being addressed at the FACET test facility at SLAC and the remaining issues may be addressed in the proposed FACET-II facility. Together these facilities will carry out the specific R&D aimed at a feasibility assessment within a decade. The promising PWFA technology may then be used for applications with strong physics interest and gradually increasing complexity in a staged approach. An informed decision about possible low energy applications could be made soon after critical issues will have been addressed in FACET-II starting to operate by 2018. Such low energy applications with possible use by 2025 would help to further develop and validate the technology by integration of the various systems and to get operational experience which is necessary before a more complex or higher energy application can be envisioned by about 2031.